\documentclass[aps,pre,floatfix,twocolumn,10pt,superscriptaddress]%
{revtex4-1}
\pdfoutput=1
\usepackage{natbib}
\usepackage{amssymb}
\usepackage{amsfonts}
\usepackage{amsmath}
\usepackage[english]{babel}
\usepackage[T1]{fontenc}
\usepackage{lmodern}

\usepackage{graphicx}

\usepackage{hyperref}
\hypersetup{%
colorlinks=true,%
citecolor=blue,%
filecolor=blue,%
linkcolor=blue,%
urlcolor=blue,%
pdfauthor={V. H. Purrello, J. L. Iguain, A. B. Kolton},%
pdftitle={Roughening of the anharmonic Larkin model}
}

\newcommand{\anhexp}{n}
\newcommand{\ampdis}{\kappa}

\begin{document}
\author{V. H. Purrello}
\email{vpurrello@ifimar-conicet.gob.ar}
\author{J. L. Iguain}
\email{iguain@mdp.edu.ar}
\affiliation{Instituto de Investigaciones F\'isicas de Mar del Plata (IFIMAR),
Facultad de Ciencias Exactas y Naturales,
Universidad Nacional de Mar del Plata,
Consejo Nacional de Investigaciones Cient\'ificas y T\'ecnicas (CONICET),
De\'an Funes 3350, B7602AYL Mar del Plata, Argentina}

\author{A. B. Kolton}
\email{koltona@cab.cnea.gov.ar}
\affiliation{Centro At\'omico Bariloche and Instituto Balseiro,
Comisi\'on Nacional de Energ\'ia At\'omica (CNEA),
Consejo Nacional de Investigaciones Cient\'ificas y T\'ecnicas (CONICET),
Universidad Nacional de Cuyo (UNCUYO),
Av. E. Bustillo 9500, R8402AGP San Carlos de Bariloche, R\'io Negro, Argentina}

\title{Roughening of the anharmonic Larkin model}

\begin{abstract}
We study the roughening of $d$-dimensional directed elastic interfaces subject
to quenched random forces. As in the Larkin model, random forces are considered
constant in the displacement direction and uncorrelated in the perpendicular
direction. The elastic energy density contains an harmonic part, proportional to
$(\partial_x u)^2$, and an anharmonic part, proportional to $(\partial_x
u)^{2n}$, where $u$ is the displacement field and $n>1$ an integer. By heuristic
scaling arguments, we obtain the global roughness exponent $\zeta$, the dynamic
exponent $z$, and the harmonic to anharmonic crossover length scale, for
arbitrary $d$ and $n$, yielding an upper critical dimension $d_c(n)=4n$. We find
a precise agreement with numerical calculations in $d=1$. For the $d=1$ case we
observe, however, an anomalous ``faceted'' scaling, with the spectral roughness
exponent $\zeta_s$ satisfying $\zeta_s > \zeta > 1$ for any finite $n>1$, hence
invalidating the usual single-exponent scaling for two-point correlation
functions, and the small gradient approximation of the elastic energy density in
the thermodynamic limit. We show that such $d=1$ case is directly related to a
family of Brownian functionals parameterized by $n$, ranging from the
random-acceleration model for $n=1$, to the L\'evy arcsine-law problem for $n =
\infty$. Our results may be experimentally relevant for describing the
roughening of nonlinear elastic interfaces in a Matheron-de Marsilly type of
random flow.
\end{abstract}


\maketitle


\section{Introduction}
\label{sec:introduction}
The general study of elastic interfaces in random media is relevant for
understanding generic properties displayed by a variety of experimental systems,
and to successfully classify them into universality
classes~\cite{Fisher1998,Kardar1998}. Disorder adds pinning, metastability, and
generates macroscopically rough interfaces, while the elasticity tends to
flatten them. When driven by a uniform force $F$, extended interfaces display a
nontrivial zero-temperature depinning transition from a static to a sliding
regime at a threshold value $F_c$~\cite{Leschhorn1997,chauve2000}, while for
finite temperatures a collective thermally activated creep motion persists below
the depinning threshold,
$F<F_c$~\cite{ioffe_creep,nattermann_creep_full,chauve2000,kolton2009}. At zero
force, the interface gets trapped in one of the many available deep metastable
states~\cite{blatter_vortex_review,nattermann2000}. In all cases, disorder
induces statistically self-affine rough geometries in the putative steady
states.

Universality allows convenient minimalistic models to capture the relevant
disorder-elasticity interplay. A rather minimalistic, yet nontrivial model, is
the one describing the position of a directed elastic interface at a time $t$ as
an univalued scalar displacement field $u(x,t)$, where $x \in \mathbb{R}^d$,
with $d$ the internal dimension of the manifold ($d=0$ a particle, $d=1$ for a
string, $d=2$ for a sheet, etc.), $D=d+1$ being the dimension of the space. Such
elastic interfaces do not allow the formation of overhangs nor pinch-off
bubbles. Specifically, we consider an elastic interface, subject to a pinning
force $g[u(x,t),x]$ with a given elastic energy $E_\text{el}(u)$. If we assume a
nondriven overdamped relaxational dynamics at zero temperature, then the
equation of motion of the interface results in
\begin{equation}
\partial_t u(x,t) = -\frac{\delta E_\text{el}}{\delta u(x,t)} + g[u(x,t),x].
\label{eq:motion}
\end{equation}
A rather generic disorder is specified by the average over disorder realizations
of the pinning force $\overline{g(x)}=0$, and its spatial autocorrelation
function $\overline{g(x,u)g(x',u')}=\ampdis^2 \delta(x-x')\Delta_{\xi}(u-u')$.
Here, $\Delta_{\xi}(u)$ is some even function of $u$ with $\xi$ denoting its
shortest characteristic length. If we choose $\Delta_{\xi}(0)=1$, then $\ampdis$
measures the strength of the disorder (see, for instance,
Ref.~\cite{chauve2000}).

The simplest harmonic form $E_\text{el} = \int dx (c_2/2)(\partial_x u)^2$, with
$c_2 \geq 0$ the elastic constant, leads to the celebrated (zero-temperature)
quenched Edwards-Wilkinson (QEW) equation~\cite{barabasi}:
\begin{equation}
\label{eq:quenched_EW}
\partial_t u(x,t) = c_2 {\partial^2_x u} + g[u(x,t),x].
\end{equation} 
The main difficulty of Eq.~(\ref{eq:quenched_EW}) is the nonlinearity and
heterogeneity of the pinning forces $g(u,x)$. This has led Larkin and coworkers
~\cite{larkin_ovchinnikov_pinning,blatter_vortex_review} to approach the problem
perturbatively in the disorder such that, at first order, the pinning force can
be simply approximated by an $x$-dependent but $u$-independent quenched random
force, $g[u(x,t),x] \approx f(x)$, with
\begin{eqnarray}
\overline{f(x)} &=& 0,\;
\overline{f(x)f(x')}=\ampdis^2 \delta(x-x').
\label{eq:pinningforceprops}
\end{eqnarray}
Replacing $g$ by $f$ in Eq.~(\ref{eq:quenched_EW}), we obtain the linear
equation
\begin{equation}
\label{eq:larkinmodel}
\partial_t u(x,t) = c_2 {\partial^2_x u} + f(x), 
\end{equation}
governing the dynamics of the so-called Larkin model (LM). Since
Eq.~(\ref{eq:larkinmodel}) is translationally invariant, bulk disorder does not
pin the interface. However, it makes the interface evolve to a rough steady
state.

In spite of its simplicity, the LM has many interesting properties and is a
fundamental model in the theory of disordered elastic systems. The linear
Eq.~(\ref{eq:larkinmodel}) is known to correctly describe the behavior of the
QEW interfaces, which evolve according to the nonlinear
Eq.~(\ref{eq:quenched_EW}), at scales smaller than the so-called Larkin length
$L_c \sim (c \xi/\ampdis)^{\frac{2}{4-d}}$, for $d < 4$.  Above $L_c$ the LM
solution crossovers to the so-called random-manifold regime. Beyond the LM
description, further crossovers at larger scales are still possible, depending
on the properties of $\Delta_{\xi}(u)$, and also on the drive
$F$~\cite{blatter_vortex_review,chauve2000,Giamarchi2002,kolton2009}. In any
case, the LM is relevant to estimate the fundamental physical units for
expressing the large-scale solution of the full QEW problem. Quite remarkably,
$L_c$ can be related to the macroscopic critical depinning threshold $F_c \sim
\ampdis/L_c^{d/2}$, the elementary pinning energy barrier $U_c = F_c L_c^{d}
\xi$, and the finite size crossover to
metastability~\cite{blatter_vortex_review,nattermann2000,chauve2000}. Above
$L_c$, in the random-manifold regime, pinning and elastic energies scale in the
same way as $\sim U_c (L/L_c)^{\theta}$, with $\theta$ a characteristic
exponent, and the interface global width as $\xi(L/L_c)^{\zeta}$, with $\zeta$ a
roughness exponent. It is worth also noting that the LM is  relevant per se if
the pinning forces have a large correlation length in the direction of interface
displacements, such that $L_c$ is large compared to system size. This can be
achieved if $\xi$ is very large, the system is elastically stiff or the disorder
very weak. Such situations can arise experimentally, displaying a finite-size
pinning crossover~\cite{dolz2010}.

Interestingly, the physics of the LM is not only relevant for describing small
length scales. Pinning forces such as $f(x)$ can be also generated via
coarse-graining at large length scales $L \gg L_c$ and dominate over other
forces in driven disordered elastic systems. This happens if the pinning force
correlator $\Delta_{\xi}(u)$ has a periodicity~\cite{ledoussal2002}. One simple
example is a QEW interface, Eq.~(\ref{eq:quenched_EW}), in a disordered medium
with periodic boundary conditions in the direction of displacements, and an
additional driving force $F$. Another less trivial example, is a one-dimensional
elastic chain of particles (or interfaces) with average separation $a_0$, driven
in a completely uncorrelated random potential~\cite{Bustingorry2010}. Such
models are often used to study friction~\cite{persson2000,cule1998}, but the
physics is also relevant for charge density waves
depinning~\cite{ledoussal2002,brazovskii2004}. In these cases, the large-scale
roughness is described by the solution of Eq.~(\ref{eq:larkinmodel}).

The LM being a fundamental model in the theory of disordered elastic systems, it
is worth analyzing how their properties change under the influence of additional
physically motivated terms in Eq.~(\ref{eq:larkinmodel}). In this paper, we
focus in the influence of anharmonic corrections to the elasticity. To motivate
the introduction of such corrections, we note that the LM, being linear, can be
easily solved for a particular realization of $f(x)$ in any dimension $d$, and
averages over realizations are straightforward. Much universal information can
be obtained by studying the critical nonstationary relaxational dynamics, from a
flat initial condition at the origin of displacements, i.e., $u(x,t=0)=0$. Since
this process is dominated by a single dynamical growing length-scale $l(t)$, for
long enough times before global equilibration.

We define the structure factor of the manifold as $S(q,t)\equiv
\overline{|\hat{u}(q,t)|^2}$, where \mbox{$\hat{u}(q,t)=  L^{-d/2} \int dx\;
u(x,t)\, e^{-i q x}$} is the Fourier transform of $u(x,t)$. Using that
$\hat{u}(q,t=0)=0$ we get,
\begin{equation}
S(q,t) \propto
  \frac{\ampdis^2}{c_2^2 q^4}\left(1-e^{-c_2q^2 t}\right)^2.
\label{eq:LMscaling}
\end{equation}
Then, the structure factor satisfies the general scaling $S(q,t)\sim
q^{-(d+2\zeta)}G(q\,t^{1/z})$, with $G(y)=y^{2(\zeta-\zeta_s)}$ for $y \gg 1$,
and $G(y)=y^{d+2 \zeta}$ for $y \ll 1$ ~\cite{PhysRevLett.84.2199}. In our case,
it is easy to check that the so-called global roughness exponent is
$\zeta=(4-d)/2$, and coincides with the spectral roughness exponent
$\zeta_s$, while the so-called dynamical exponent, related to the growing
length-scale $l(t) \sim t^{1/z}$, is $z=2$. In the steady state, roughly
reached at times $t$ such that $l(t) \sim L$, $S(q,t\to \infty)\sim
q^{-(d+2\zeta)}$. For an interface of size $L$, this gives the global squared
width with respect to the center of mass position $W^2 \equiv
\overline{[u(x,t)-v_{\text{cm}}t]^2}=\int dq\;S(q,t) \sim L^{2\zeta}$, with
$v_{\text{cm}} = L^{-d} \int dx f(x) \sim L^{-d/2}$ the finite-size residual
center of mass velocity. The interface then becomes macroscopically self-affine,
with exponent $\zeta$. That is, the rescaling $x' \to b\,x$ and $u' \to
b^\zeta\,u$ leads to a statistically equivalent interface.

Interestingly, in the $d=1$ LM, we have $\zeta_s=\zeta>1$. This situation, which
has been called the super-rough case in Ref.~\cite{PhysRevLett.84.2199},
has a physical peculiarity. It implies that the harmonic elastic approximation
in Eq.~(\ref{eq:larkinmodel}) to the local elastic couplings must break down in
the thermodynamic limit, since $W/L \sim L^{\zeta-1} \to \infty$ as $L\to
\infty$. Local elongations are thus not  bounded in the thermodynamic limit. A
similar situation occurs for the roughness exponent at the depinning threshold
for the driven QEW model, where $\zeta \approx
1.25$~\cite{depinninginvariousd,Ferrero_2013}. To remedy this situation, in
Ref.~\cite{Rosso2001b} ad hoc anharmonic corrections to the elastic energy
were introduced, such that
\begin{equation}
E_\text{el}(u) = \int dx\;
  \left[\frac{c_2}{2}\,(\partial_x u)^2 +
        \frac{c_{2\anhexp}}{2\anhexp}\,(\partial_x u)^{2\anhexp}\right],
\label{eq:nonharmonicelasticity}
\end{equation}
with $\anhexp=2,3,4...$, and constants $c_2>0$, $c_{2\anhexp}>0$. It is worth
noting that this kind of correction is Hamiltonian, convex, and being a
correction to the elasticity only, translational invariant. In particular, note
that the presence of the anharmonic correction breaks the tilt-symmetry of the
full QEW equation, since both $g(u,x)$ and the harmonic elasticity, in contrast
with the $(\partial_x u)^{2\anhexp}$ term for $n>1$, are statistically invariant
by the transformation $u \to u-s x_\alpha$, with $\alpha$ denoting any of the
$d$ internal directions and $s$ the parameter measuring the tilt deformation.

For large $\anhexp$, Eq.~(\ref{eq:nonharmonicelasticity}) is equivalent to
impose a hard-constraint to local elongations, an usual modeling of directed
polymers~\cite{halpinhealey1995}. The proposed nonquadratic term succeeds to
save the elastic approximation in the thermodynamic limit at the depinning
transition of the QEW model, with a new (``physical'') roughness exponent $\zeta
= \zeta_s \approx 0.63$ for all $\anhexp>1$~\cite{Rosso2001b,kolton2009}. Quite
remarkably, this value is in perfect agreement with the roughness at the
depinning transition of the paradigmatic quenched Kardar-Parisi-Zhang (QKPZ)
model~\cite{barabasi}, which is inherently a nonequilibrium effective equation
that cannot be derived directly from a Hamiltonian or free energy. The
anharmonic model hence allows us to study a model with a Hamiltonian, and a
well-defined equilibrium state at $F=0$, which nevertheless spontaneously
generates the ubiquitous Kardar-Parisi-Zhang (KPZ)~\cite{kpz1986} term, when it
is driven by a force $F$. Thus, one may ask whether the breakdown of the small
gradient approximation in the elastic energy of the $d=1$ Larkin model
[Eq.~(\ref{eq:larkinmodel})] is, similarly, protected by introducing a nonlinear
elasticity of the form proposed in Eq.~(\ref{eq:nonharmonicelasticity}), and how
$\zeta$ and $\zeta_s$ would change upon its addition. Solving the resulting
``anharmonic Larkin model'' would also allow us to find a possibly modified
Larkin length $L_c$ and related quantities, which are fundamental to estimate
both the critical force and the crossover length to the random manifold regime,
for the anharmonic depinning model defined by Eqs.~(\ref{eq:motion})
and~(\ref{eq:nonharmonicelasticity}). The depinning transition of such model was
analyzed in Ref.~\cite{Rosso2001b}.

Motivated by the above phenomenology, in this work, we study the Larkin model
with anharmonic elasticity, for a general dimension $d$ and $\anhexp \geq 2$. We
obtain the global roughness exponent as a function of $\anhexp$ and $d$, and
also describe the crossover from harmonic to anharmonic regimes of roughness,
when two terms, one with $\anhexp=1$ and another with $\anhexp>1$ coexist. For
the special $d=1$ case, where the small elastic deformation approximation is
compromised, we show that, unlike the depinning model, the anharmonic correction
is \textit{never} able to reduce $\zeta$ below unity, even in the large
$\anhexp$ almost hard-constraint limit. Moreover, we show that for all
$\anhexp>1$ the $d=1$ interface displays anomalous scaling properties, with
$\zeta_s > \zeta \geq 1$, the so-called ``faceted regime'' in
Ref.~\cite{PhysRevLett.84.2199}. Interestingly, we show also that this
anomalous case is closely connected to an $\anhexp$-parameterized family of
Brownian functionals which interpolate between the random-acceleration process
for $\anhexp=1$, to the arcsine law L\'{e}vy stochastic process, for
$\anhexp=\infty$. For $d>1$, however, we find $\zeta \leq 1$.

This article is organized as follows: in Sec.~\ref{sec:model} we describe the
anharmonic Larkin model, the observables of interest, and the methods. In
Sec.~\ref{sec:scalingarguments} we derive, via heuristic arguments, the global
roughness and dynamical exponents, and also the harmonic-anharmonic crossover
length. In Sec.~\ref{sec:numerics}, we numerically test these predictions for
$d=1$, solving both the relaxational dynamics of an interface as a function of
time, and the statics, for two different boundary conditions. In
Sec.~\ref{sec:discussions}, we discuss the relation for the anomalous $d=1$ case
with a family of Brownian functionals. Finally, in Sec.~\ref{sec:conclusion} we
present our conclusions along with new open questions, and we suggest some
possible applications for our results.

\section{Model and Observables}
\label{sec:model}
\begin{figure}[tb]
\includegraphics[width=0.5\textwidth]{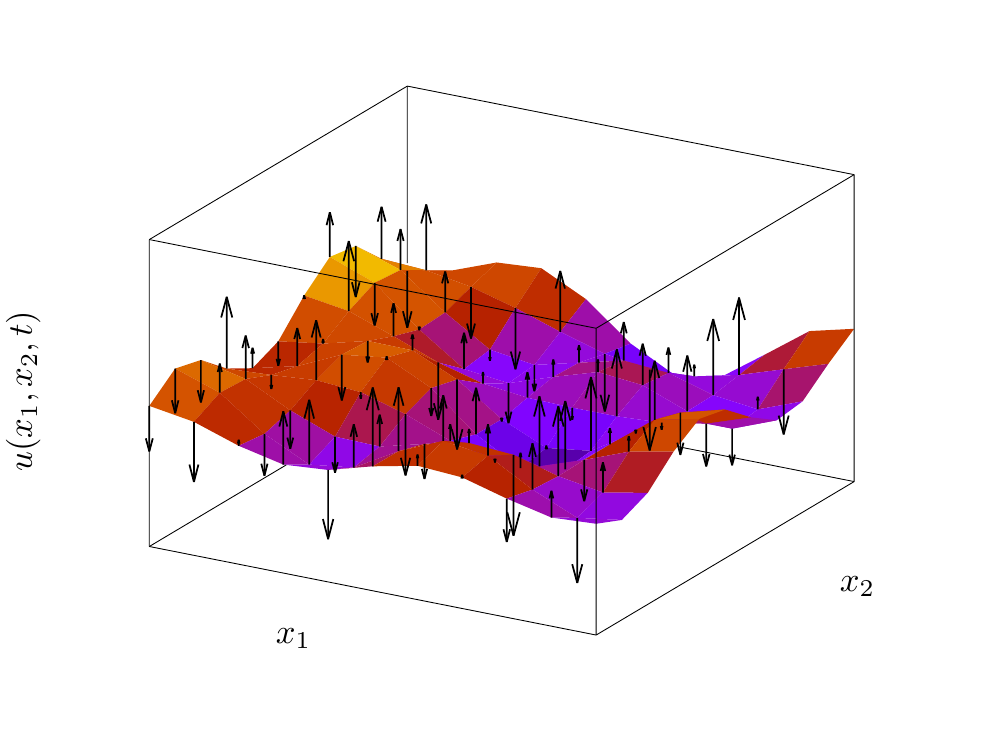}
\caption{Schematics of the anharmonic Larkin model (ALM) for the particular
$d=2$ case, corresponding to a two dimensional interface in a three-dimensional
medium. The directed interface is described by an instantaneous scalar
displacement field in the vertical direction $u(x,t)$ governed by
Eq.~(\ref{eq:general_differential}), with $x \equiv (x_1,x_2)$ the internal
coordinate. The elasticity of the interface is nonlinear, with an elastic
energy given by Eq.~(\ref{eq:nonharmonicelasticity}). Arrows indicate that each
point $x$ of the interface is subject to a quenched uncorrelated scalar random
force $f(x)$, described by Eq.~(\ref{eq:pinningforceprops}).}
\label{fig:esquemaalm}
\end{figure}

We consider the anharmonic Larkin model using Eq.~(\ref{eq:motion}) in the
Larkin approximation $g(u,x) \to f(x)$, and Eq.~(\ref{eq:nonharmonicelasticity})
for the elastic energy. The resulting equation of motion reads
\begin{equation}
\partial_t u(x,t) =
c_2 \partial^2_x u + c_{2\anhexp} \partial_x 
\left[\left(\partial_x u\right)^{2\anhexp-1}\right] + f(x),
\label{eq:general_differential}
\end{equation}
where $\anhexp>1$. We will call Eq.~(\ref{eq:general_differential}) the
``anharmonic Larkin model'' (ALM). Fig.~\ref{fig:esquemaalm} schematically
represents the model for the particular $d=2$ case.

Let us consider an interface of linear dimension $L$. The time-dependent global
width for the relaxation process $W(L,t)$ is defined by the mean-squared
fluctuations around the center of mass,
\begin{equation}
W^2(L,t) = \overline{\frac{1}{L^d} \int dx\; (u-v_{\text{cm}}t)^2},
\label{eq:w2_def}
\end{equation}
where \mbox{$v_{\text{cm}} = L^{-d} \int dx\;f(x) \sim L^{-d/2}$} [from
Eq.~(\ref{eq:general_differential})] is the finite-size disorder-dependent but
constant center of mass velocity, that may be observed for periodic or free
boundary conditions. It is also useful to compute the time-dependent structure
factor $S(q,t,L)$, as the interface displacement spatial power spectrum,
\begin{equation}
S(q,t,L) = \overline{|\hat{u}(q,t)|^2}, 
\end{equation}
with $\hat{u}(q,t)$ the Fourier transform of $u(x,t)$ as defined just before
Eq.~(\ref{eq:LMscaling}).

Solving Eq.~(\ref{eq:general_differential}) from a flat initial condition
\mbox{$u(x,t=0)=0$} allows us to extract several universal exponents
characterizing the generically expected critical relaxational dynamics. Such
dynamics are expected to be controlled by a single growing correlation length
scale,
\begin{equation}
l(t) \sim t^{1/z},
\end{equation}
with $z$ defining the dynamical exponent. Such scaling is expected to be valid
at times larger than the microscopic time scale, but smaller than the time at
which the system becomes completely correlated, i.e., when $l(t) \sim L$.
However, from $W^2(L,t)$, and $S(q,t,L)$ we can define in principle three
different roughness exponents ($z$, $\zeta$, and $\zeta_s$) characterizing the
random geometry below $l(t)$~\cite{PhysRevLett.84.2199}.

The global roughness exponent $\zeta$ is defined from
\begin{equation}
W(t,L) \sim t^{\zeta/z} \tilde{W}[L/l(t)],
\label{eq:Wscaling}
\end{equation}
with
\begin{equation}
\tilde{W}(x) \sim
  \begin{cases}
    x^\zeta & \text{ if } x \ll 1 \\
    \text{const} & \text{ if } x \gg 1,
  \end{cases}
\end{equation}
or more directly from the stationary interface width
\begin{equation}
W(t\to \infty,L) \sim L^{\zeta}. 
\label{eq:WvsLscaling}
\end{equation}

Both the global $\zeta$ and the spectral $\zeta_s$ roughness exponents can be
obtained from the expected scaling
\begin{equation}
S(q,t) = q^{-(2\zeta+1)}\ \tilde{S}\left({qt^{1/z}}\right),
\label{eq:sq_scaling}
\end{equation}
where $\tilde{S}(x)$ is the scaling function and has the general form
\begin{equation}
\tilde{S}(x) \sim
  \begin{cases}
    x^{2(\zeta-\zeta_s)}, &\; \text{ if } x \gg 1 \\
    x^{2\zeta+1}, &\; \text{ if } x \ll 1.
  \end{cases}
\label{eq:spectraldefinition}
\end{equation}
The stationary limit is reached when the correlation length $l(t)$ reaches $L$,
at times of order $L^z$. Thus,
\begin{equation}
S(q) \sim q^{-(2\zeta_s+1)} L^{2(\zeta-\zeta_s)}.
\label{eq:steadysofq}
\end{equation}

At zero temperature, the nonstationary scaling behavior from
Eqs.~(\ref{eq:Wscaling}), (\ref{eq:WvsLscaling}), (\ref{eq:sq_scaling}), and
(\ref{eq:spectraldefinition}), and also the stationary scaling described by
Eq.~(\ref{eq:steadysofq}), are verified by the QEW [Eq.~(\ref{eq:quenched_EW})]
at or above the depinning threshold $F \geq F_c$~\cite{kolton2006b,Kolton2009b},
and in particular by the harmonic Larkin model [Eq.~(\ref{eq:larkinmodel})] for
any force, as shown in Eq.~(\ref{eq:LMscaling}). In all these cases, a single
growing correlation length controls the relaxation toward a unique self-affine
stationary state, without memory of the initial condition. Different roughness
exponents are obtained in each case. For instance, the $d=1$ QEW equation at
$F_c$ has $\zeta =  \zeta_s \approx 1.25$ and $z \approx
1.433$~\cite{Ferrero_2013}, while at $F>F_c$, it crossovers to the
Edwards-Wilkinson exponents $\zeta = \zeta_s = 1/2$ and $z=2$. The harmonic
Larkin model, however, as discussed in Sec.~\ref{sec:introduction}, has
$\zeta=\zeta_s=(4-d)/2$ and $z=2$. In the following sections, we will show that
the ALM also displays the same scaling forms, and we will obtain their exponents
$\zeta$, $\zeta_s$, and $z$ for $\anhexp=1,2,...,\infty$.

\section{Scaling Arguments}
\label{sec:scalingarguments}
The dynamical and roughness exponents of the EW, the LM and other linear models
can be successfully obtained by simple scaling arguments (see
Ref.~\cite{barabasi} for many more relevant examples). Although such approach
may fail in general (for instance, KPZ, QKPZ, or QEW equations require
renormalization group calculations instead), we will employ the same naive
methodology for the ALM nonlinear dynamics in
Eq.~(\ref{eq:general_differential}). For simplicity, at first we consider only
the nonlinear elastic term. Putting $c_2=0$, Eq.~(\ref{eq:general_differential})
reduces to
\begin{equation}
\partial_t u(x,t) = c_{2\anhexp} \partial_x
  \left[\left(\partial_x u\right)^{2\anhexp-1}\right] + f(x).
\label{eq:anharmonic}
\end{equation}

Hence, to obtain the scaling exponents, we follow the standard naive procedure,
which consists in rescaling  space and time as
\begin{align}
  x &\rightarrow x' \equiv b\,x,\\
  u &\rightarrow u' \equiv b^\zeta\,u,\\
  t &\rightarrow t' \equiv b^z\,t.
\end{align}
Then, making the strong assumption that the elasticity and pinning parameters
\textit{are not} changed by rescaling, after this transformation,
Eq.~(\ref{eq:anharmonic}) results in
\begin{multline}
b^{\zeta-z} \partial_t u =\\
  b^{(\zeta-1)(2\anhexp-1)-1} c_{2\anhexp} \partial_x
    \left[\left(\partial_x u\right)^{2\anhexp-1}\right] + b^{-d/2} f,
\end{multline}
where we have used that the random uncorrelated pinning forces scale as a
$d$-dimensional random walk.

We finally require the resulting equation must be invariant under the
transformation, which leads to the Flory-like exponents,
\begin{align}
\zeta(d,\anhexp) &= \frac{4\anhexp-d}{4\anhexp-2}, \label{eq:prediction_zeta}\\
  z(d,\anhexp) &= \zeta(d,\anhexp) + \frac{d}{2},
\label{eq:prediction_z}
\end{align}
for $\anhexp=1,2,...,\infty$, and general dimension $d$. 

As expected, for $\anhexp=1$, we recover the LM exponents, $\zeta=(4-d)/2$ and
$z=2$, and the known upper critical dimension $d_c(\anhexp=1)=4$. From the
prediction in Eqs.~(\ref{eq:prediction_zeta}) and~(\ref{eq:prediction_z}), it is
worth noting the following:
\begin{itemize}
\item[(1)] The upper critical dimension increases with $\anhexp$, as
$d_c(\anhexp)=4\anhexp$.
\item[(2)] In the hard-constraint case,  corresponding to \mbox{$\anhexp \to
\infty$}, we get, for a fixed dimension $d$, that $\zeta \to 1$, and $z \to
1+d/2$.
\item[(3)] The anharmonic elastic energy density scales as
$L^{2\anhexp[\zeta(d,\anhexp)-1]}$, thus for $\anhexp>1$ it is thermodynamically
bounded only for $d>2$, since $\zeta(d > 2,\anhexp) < 1$. The $d=2$ case is
always marginal, $\zeta(d=2,\anhexp) = 1$, while the $d<2$ case diverges, since
$\zeta(d < 2,\anhexp) > 1$.
\item[(4)] The spectral roughness exponent $\zeta_s$ does not come out from the
arguments.
\end{itemize}

If the predicted exponents are valid, then we arrive to the striking conclusion
that the unbounded local displacements predicted for the $d=1$ LM remain
unbounded, in spite of the anharmonic elasticity, for any value of $\anhexp$.
Therefore, the anharmonic correction to the elasticity can not save the small
local deformation assumption behind the gradient expansion of the elastic energy
density, even in the $n \to \infty$ hard-constraint limit, in the large-size
limit. This is in sharp contrast to what happens for the $d=1$ QEW equation at
depinning, where the elastic approximation is saved by the very same correction,
making the depinning roughness exponent to change from $\zeta_{\text{dep}}
\approx 1.25$ to the $n$-independent ($n>1$) value $\zeta_{\text{dep}}\approx
0.63$~\cite{Rosso2001b}~\footnote{Note that if the ALM is considered as a
short-scale approximation of the QEW with anharmonic
corrections~\cite{Rosso2001b}, it applies only below the corresponding finite
Larkin length, calculated in the Appendix. In this case, the elastic
approximation is not necessarily compromised in the large scale limit in spite
of the super-rough (anharmonic) Larkin regime.}.

Let us now consider both the usual harmonic elasticity term and the nonlinear
correction together, i.e., $c_2 > 0$ and $c_{2\anhexp} > 0$, for a given
$\anhexp>1$, in Eq.~(\ref{eq:general_differential}). In this case, the harmonic
term should dominate the behavior for small distortions, so there might be an
\textit{anharmonic crossover} length $L_\text{anh}$ between two regimes of
roughness, from the harmonic LM ($c_2>0$, $c_{2\anhexp}=0$) to the previously
analyzed purely anharmonic ALM ($c_2=0$, $c_{2\anhexp}>0$) universality classes.
Heuristically, we can propose that, for a length-scale $l>L_\text{anh}$, the
purely anharmonic ALM displacement behaves as $u \sim u_\text{anh}
(l/L_\text{anh})^{\zeta(d,\anhexp)}$; while for $l<L_\text{anh}$, we have the
harmonic result $u \sim (\ampdis/c_{2}) l^{\zeta(d,\anhexp=1)} = (\ampdis/c_{2})
l^{(4-d)/2}$. Sharply matching these two regimes at $L_\text{anh}$ implies
\begin{equation}
u_\text{anh}=\frac{\ampdis}{c_{2}}\,L_\text{anh}^{(4-d)/2}.
\end{equation}
We propose that the crossover occurs when the two elastic
energy terms are equally important, $c_{2}
(u_\text{anh}/L_\text{anh})^2=(c_{2\anhexp}/\anhexp)
(u_\text{anh}/L_\text{anh})^{2\anhexp}$, thus,
\begin{equation}
u_\text{anh} =
    \left({\frac{\anhexp c_2}{c_{2\anhexp}}}\right)^{\frac{1}{2(\anhexp-1)}}\,
  L_\text{anh}
             = \frac{\ampdis}{c_{2}}\,L_\text{anh}^{(4-d)/2},
\label{eq:energy_balance}
\end{equation}
which leads to the crossover length,
\begin{equation}
 L_\text{anh} = \left({\frac{c_2}{\ampdis}}\right)^\frac{2}{2-d}\,
  \left({\frac{\anhexp c_2}{c_{2\anhexp}}}\right)^{\frac{1}{(2-d)(\anhexp-1)}}.
 \label{eq:prediction_lanh}
\end{equation}
This means that, for  the combined elasticities, we would have, in the
stationary limit of a large system $L \gg L_{\text{anh}}$, a crossover behavior
as a function of the length-scale in the correlation functions. For instance,
the mean-square width [Eq.~(\ref{eq:w2_def})] is expected to satisfy
\begin{equation}
W^2(L,t\to \infty) = L^{2\zeta(d,n=1)}\,w(L/L_{\text{anh}}),
\end{equation}
with $w(x)=x^{2[\zeta(d,n>1)-\zeta(d,n=1)]}$, for $x \gg 1$; and \mbox{$w(x)
\sim \text{constant}$}, for $x \ll 1$.

In the following two sections, we numerically check  all these heuristic scaling
predictions, for the special $d=1$ case.

\section{Numerical Results for $d=1$}
\label{sec:numerics}
To test the validity and robustness of the scaling predictions, we have found
convenient to analyze separately the static solution, using appropriate boundary
conditions. With these numerical methods, we are able to access all the
exponents $\zeta$, $\zeta_s$, and $z$, and the crossover length
$L_{\text{anh}}$.

\subsection{Static solution}
\label{sub:statics}
We study the static geometric properties of the ALM,  by solving
Eq.~(\ref{eq:general_differential}) in the stationary limit, where the elastic
and
random forces exactly balance, i.e.,
\begin{equation}
c_2 \partial^2_x u + c_{2\anhexp} \partial_x [(\partial_x u)^{2\anhexp-1}] =
  -f(x).
\label{eq:gralstaticeq}
\end{equation}

\subsubsection{Global roughness exponent}
Since the anharmonic term dominates the geometry at large length scales, we
first focus on the purely anharmonic ALM by fixing $c_2=0$ in
Eq.~(\ref{eq:gralstaticeq}). This is then straightforward to solve:
\begin{equation}
u(x) =  -\int_0^x dx'
        \left[\int_0^{x'}dx''\frac{f(x'')}{c_{2\anhexp}}-(\partial_x
              u(0))^{2\anhexp-1}\right]^{\frac{1}{2\anhexp-1}},
\label{eq:staticequation}
\end{equation}
where we have used the particular boundary condition {$u(0,t)=0$}. As elastic
interactions are short ranged, this particular boundary condition does not alter
the large-scale scaling properties we are interested in.

To numerically evaluate Eq.~(\ref{eq:staticequation}), we simply discretize the
$d=1$ interface in segments of size $\delta x$, and perform the sums
\begin{equation}
u_i = -\sum_{j=0}^{i-1}
\left[{\sum_{k=0}^j \frac{{f}_{k}}{c_{2\anhexp}}}\right]^{\frac{1}{2\anhexp-1}},
\;\;\; \text{for}\; i>0,
\label{eq:conf_JLI}
\end{equation}
with $u_0=0$ fixed, and $f_0 \equiv -\sum_{i=1}^L f_i$; taking care of the
statics of the whole polymer of size $L$ in the discrete description, and
accounting for the term \mbox{$[\partial_x u(0)]^{2\anhexp-1} \sim
(u_1-u_0)^{2\anhexp-1} = -f_0/c_{2\anhexp}$} in the continuum
Eq.~(\ref{eq:staticequation})~\footnote{Note that the constraint force $f_0$,
needed to fix the boundary conditions in the first displacement $u_0$ of the
polymer, is not described by the bulk random forces, with the statistical
properties in Eq.~(\ref{eq:pinningforceprops}).}. Note that, without loss of
generality, we have taken $\delta x=1$ exploiting that the pinning force is
completely uncorrelated in the internal coordinate. We have drawn $f_k$ from a
uniform distribution with zero mean and variance {$\ampdis^2 = 1/12$}.

Equation~(\ref{eq:conf_JLI}) can  then be solved very efficiently for large
system sizes, by using parallel random number generators and parallel prefix-sum
algorithms implemented in massively parallel coprocessors, such as graphic
cards. As the same scaling analysis leading to the prediction in
Eq.~(\ref{eq:prediction_zeta}) applies to Eq.~(\ref{eq:staticequation}) or
Eq.~(\ref{eq:conf_JLI}), regardless of the  fixed-end condition $u_0=0$, we
expect to get $\zeta(d=1,\anhexp)$ from the numerical evaluation of
Eq.~(\ref{eq:conf_JLI}). Since the system is not translational invariant along
$x$, it is convenient to directly measure the last-point scaling, for which we
expect $\overline{u_L^2} \sim L^{2\zeta(d=1,\anhexp)}$, for large enough $L$.

\begin{figure}[tb]
\includegraphics[width=\columnwidth]{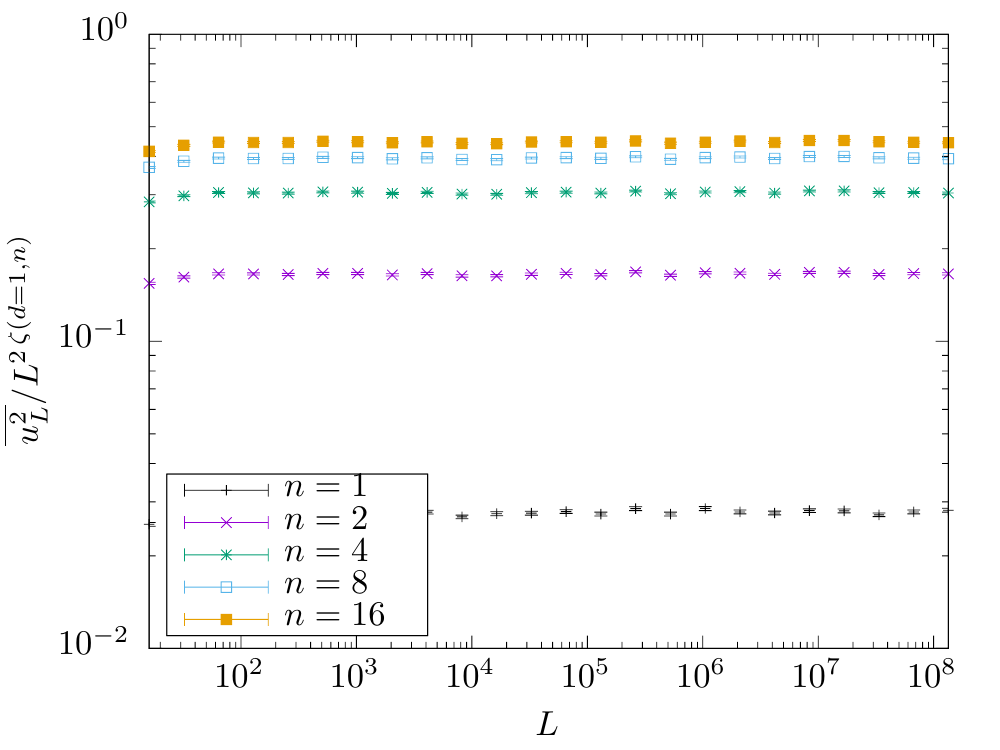}
\caption{Disorder-averaged last-end-point squared displacement of a $d=1$ ALM
interface divided by the expected size scaling, as a function of the system size
$L$, using $10^4$ samples for each value of $n$ and $L$. Constant behavior at
large $L$ confirms the validity of the predicted global roughness exponent
$\zeta(d=1,\anhexp)$ of Eq.~(\ref{eq:prediction_zeta}). The error bars are
estimated from the standard deviations.}
\label{fig:w2_vs_L}
\end{figure}

In Fig.~\ref{fig:w2_vs_L} we show there is an excellent agreement between
predicted behavior and the obtained results for various values of $\anhexp$, for
system sizes larger than around $10^2$, and up to $10^8$ string elements.

\subsubsection{Harmonic to anharmonic crossover}
To study the crossover between the LM and the ALM, we now consider the static
condition Eq.~(\ref{eq:gralstaticeq}), with $c_2 \neq 0$ and $c_{2\anhexp}\neq
0$, i.e., by including both the harmonic and the anharmonic terms. Integrating
it, we get
\begin{eqnarray}
c_2\ \partial_x u + c_{2\anhexp}\ (\partial_x u)^{2\anhexp-1} &=&
  c_2\ \partial_x u(0) + c_{2\anhexp}\ [\partial_x u(0)]^{2\anhexp-1}
    \nonumber \\
  &-& \int_0^x  dx''\;f(x'') ,
\label{eq:polynomialgradient}
\end{eqnarray}
which is an odd $(2\anhexp - 1)$-degree polynomial in $\partial_x u$. Since
$c_2>0$ and $c_{2\anhexp}>0$, it has only one real root for each $x$. By
integrating this
$x$-dependent root we obtain $u(x)$. We perform this task numerically, for the
discretized
interface $u_i$, as before; using that $c_2\ \partial_x u(0) + c_{2\anhexp}\
[\partial_x u(0)]^{2\anhexp-1} \sim -f_0$ in this mixed case. Finally, to test
the prediction for $L_\text{anh}$ [Eq.~(\ref{eq:prediction_lanh})], we use the
expected crossover scaling for the polymer end-point,
\begin{equation}
\overline{u_L^2} \sim L_{\text{anh}}^{4-d} \;h(L/L_\text{anh}),
\label{eq:anhcrossoverscaling}
\end{equation}
with $h(x)\sim x^{4-d}$, for $x \ll 1$; and $h(x) \sim x^{2\zeta(d=1,\anhexp)}$,
for $x \gg 1$.

\begin{figure}[tb]
\includegraphics[width=\columnwidth]{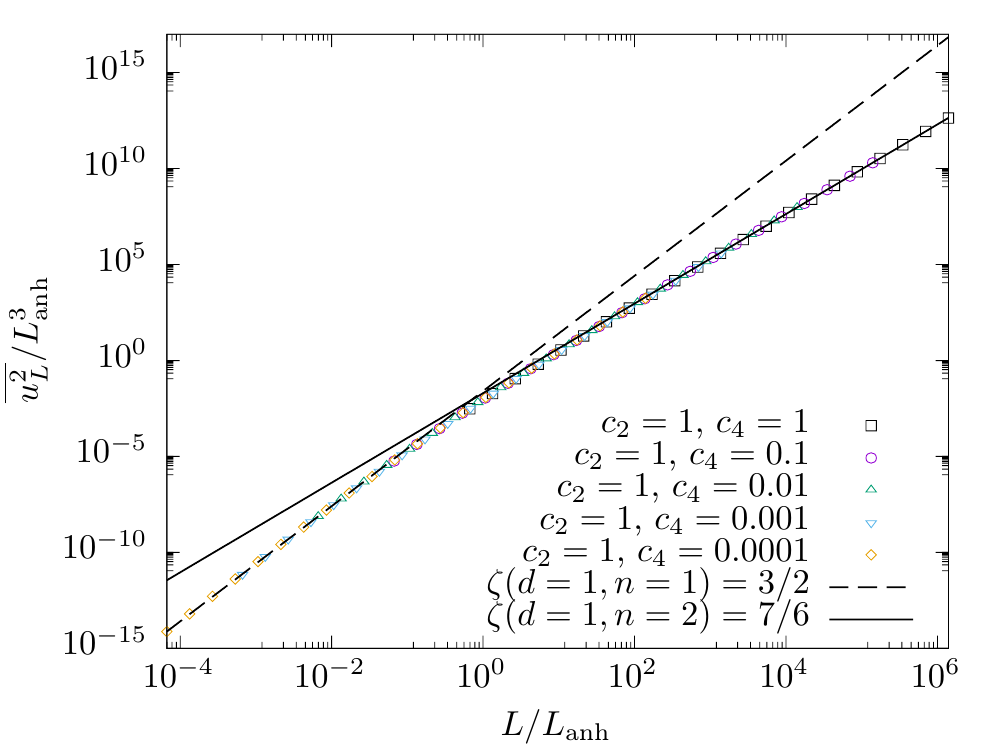}
\caption{Rescaled disorder-averaged quadratic displacement of the end-point
combining both an harmonic and an anharmonic  elasticity. The master curve
confirms the general crossover scaling predicted in
Eqs.~(\ref{eq:prediction_lanh}) and~(\ref{eq:anhcrossoverscaling}), with
$L_\text{anh}$
particularized for $d=1$ and $\anhexp=2$ [Eq.~(\ref{eq:lanh_d1})]. The dashed
and
solid lines indicate the scalings $\overline{u^2_L}\sim L^{\zeta(d=1,n=1)}$ and
$\overline{u^2_L}\sim L^{\zeta(d=1,n=2)}$, respectively, with the roughness
exponents predicted by Eq.~(\ref{eq:prediction_zeta}).}
\label{fig:w2_elast_comb}
\end{figure}

We have tested the prediction for $L_\text{anh}$ from
Eq.~(\ref{eq:prediction_lanh}), for the particular $d=1$ and $\anhexp=2$ case,
where Eq.~(\ref{eq:polynomialgradient}) becomes an analytically solvable cubic
polynomial. The prediction of Eq.~(\ref{eq:prediction_lanh}) for this case is
\begin{equation}
L_\text{anh} = \left({\frac{c_2}{\ampdis}}\right)^2\,\frac{2 c_2}{c_4}
  = 2 c_2^2 \ampdis^{-2} c_4^{-1}.
\label{eq:lanh_d1}
\end{equation}
In Fig.~\ref{fig:w2_elast_comb}, we show the results for the rescaled families
of curves, $\overline{u_L^2}/L_\text{anh}^{3}$ versus $L/L_\text{anh}$,
combining $c_2 = 1$  with various values for $c_4$, from $c_4 = 1$ to $c_4=
0.0001$. The master curve strongly supports the validity of
Eq.~(\ref{eq:prediction_lanh}).

\subsection{Dynamic Solution}
\label{sub:dynamics}
To test dynamical scaling (i.e., involving the time variable), we have performed
numerical simulations of Eq.~(\ref{eq:general_differential}) in $d=1$ with
periodic boundary conditions, and we have averaged the results over many
disorder realizations. We have implemented this by using a spatial finite
difference scheme, and we have solved the resulting system of equations
following the standard Euler method. If the discretization is $\delta x$, such
that $u_i \equiv u(x=i\delta x)$, for $i=1,2,\dots,L$, then
Eq.~(\ref{eq:general_differential}) can be approximated by
\begin{align}
\partial_t u_i &=
  \frac{c_{2}}{(\delta x)^{3}} \left(u_{i+1}+u_{i-1}-2 u_i\right) \nonumber \\
&+ \frac{c_{2\anhexp}}{(\delta x)^{2\anhexp+1}}\left[
  \left(u_{i+1}-u_{i}\right)^{2\anhexp-1} -
  \left(u_{i}-u_{i-1}\right)^{2\anhexp-1}\right] \nonumber \\
&+ f_{i},
\label{eq:implemented_discretization}
\end{align}
with $u_0 \equiv u_L$ and $u_{L+1}\equiv u_1$. The $d$-dimensional
generalization is straightforward. We draw the random forces $f_i$ from a
uniform distribution with zero mean and variance {$\ampdis^2 = 1/12$}, and we
solve Eq.~(\ref{eq:implemented_discretization}) starting from a flat
configuration
$u_i(t=0) = 0$.

\subsubsection{Time dependent structure factor}
The structure factor $S(q,t,L)$ allows us to obtain $\zeta$, and in particular,
the dynamical exponent $z$ and the spectral roughness exponent $\zeta_s$, from
the expected scaling of Eq.~(\ref{eq:sq_scaling}).

\begin{figure}[tb]
\includegraphics[width=\columnwidth]{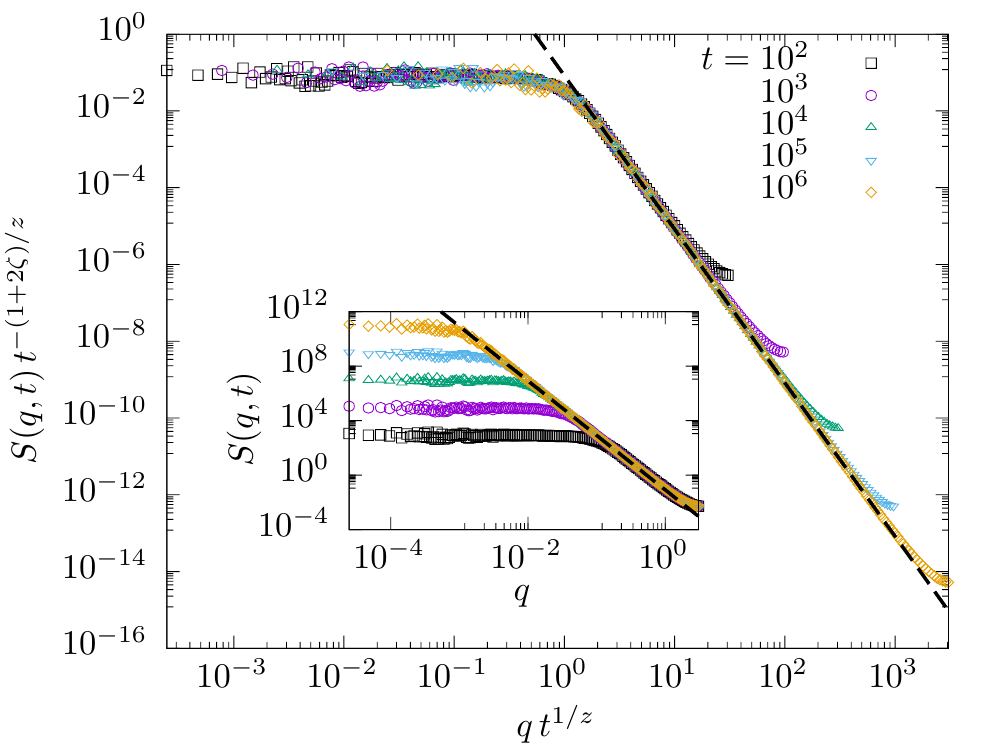}
\caption{Rescaled structure factors of the  harmonic string (i.e. $d=1$,
$\anhexp=1$) Larkin model (LM) for different times $t$, rescaled according to
Eq.~(\ref{eq:LMscaling}) with $\zeta \equiv \zeta(d=1,n=1) = 3/2$ and $z \equiv
z(d=1,n=1) = 2$. The dashed line indicates a $q^{-(1+2\zeta_s)}$ scaling, with
$\zeta_s=\zeta$. Inset: raw data.}
\label{fig:sq_scaled_alpha_1}
\end{figure}

In Fig.~\ref{fig:sq_scaled_alpha_1} we show the corresponding $S(q,t)$ for $c_2
= 1$, $c_{2\anhexp}=0$, $L=262144$, and different times $t$. The main panel
shows the raw data (displayed in the inset) rescaled using
Eq.~(\ref{eq:sq_scaling}). The result is perfectly consistent with the scaling
and confirms the exponents $\zeta=3/2$ and $z=2$ of the analytical solution of
Eqs.~(\ref{eq:prediction_zeta}) and~(\ref{eq:prediction_z}) for $d=1$ and
$\anhexp=1$.

\begin{figure}[tb]
\includegraphics[width=\columnwidth]{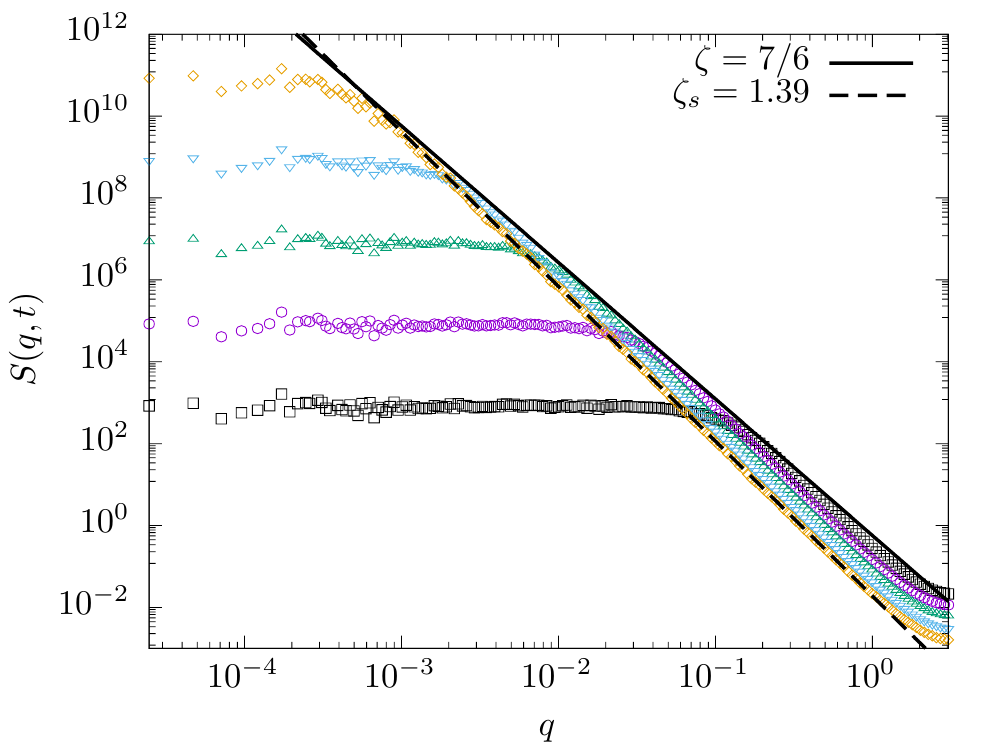}
\caption{Structure factor time evolution for the anharmonic string $\anhexp=2$,
$d=1$. The spectral roughness exponent $\zeta_s$ fits individual curves as
$S(q,t) \sim q^{-(1+2\zeta_s)}$ for $qt^{1/z} \gg 1$, while the global exponent
$\zeta$ fits the curves envelope as $S(q,t) \sim q^{-(1+2\zeta)}$. Scaling is
anomalous, $\zeta_s \neq \zeta$, in contrast with the harmonic model (see
Fig.~\ref{fig:sq_scaled_alpha_1}).}
\label{fig:sq_envolvente_alpha_2}
\end{figure}

\begin{figure}[tb]
\includegraphics[width=\columnwidth]{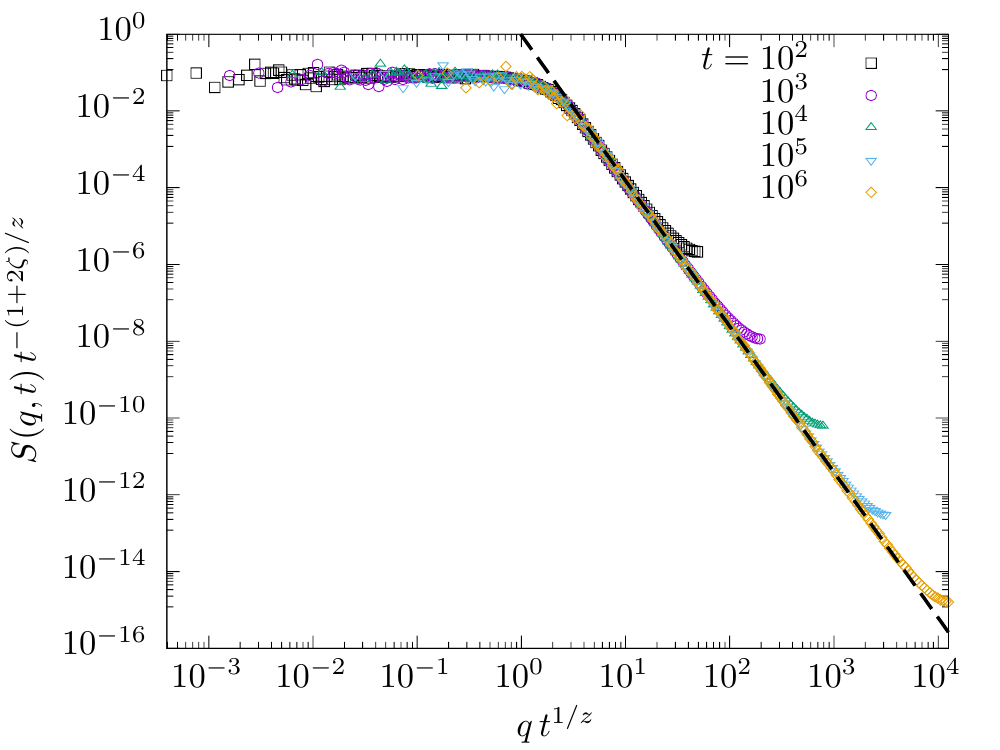}
\caption{Rescaled structure factor evolution for the $d=1$ and $\anhexp=2$
anharmonic Larkin model, using $\zeta \equiv \zeta(d=1,n=2) =7/6$ and $z \equiv
z(d=1,n=2)=5/3$, as predicted by Eqs.~(\ref{eq:prediction_zeta}) and
~(\ref{eq:prediction_z}). The dashed line corresponds to the power-law
$1/q^{1+2\zeta_s}$, indicating a spectral roughness exponent $\zeta_s \approx
1.39$.}
\label{fig:sq_scaled_alpha_2}
\end{figure}

To contrast with the pure harmonic case considered before, we now consider the
purely anharmonic case with $c_2=0$ and $c_{2\anhexp}=1$ using $\anhexp=2$. In
Fig.~\ref{fig:sq_envolvente_alpha_2}, we show the corresponding $S(q,t)$ for
different times. Interestingly, we can observe that large $q$ behavior does not
follow a master curve, but there is a downward drift by increasing times, which
differs from the harmonic result shown in the inset of
Fig.~\ref{fig:sq_scaled_alpha_1}. Moreover, fitting the large $q$ power-law
behavior for a long time $t$, yields a different power-law exponent than fitting
the power-law envelope for all the times. In Fig.~\ref{fig:sq_scaled_alpha_2} we
show that, nevertheless, the scaling of Eq.~(\ref{eq:sq_scaling}) works
perfectly
by using the global roughness exponent $\zeta(d=1,\anhexp=2) = 7/6$ and the
dynamical exponent $z(d=1,\anhexp=2) =5/3$, both predicted by dimensional
analysis. However, in spite of the perfect collapse using the predictions of the
scaling arguments of Eqs.~(\ref{eq:prediction_zeta}) and
~(\ref{eq:prediction_z})
for $d=1$ and $\anhexp=2$, the master curve displays a clearly larger and well
defined roughness exponent. From Eqs.~(\ref{eq:sq_scaling}) and
~(\ref{eq:spectraldefinition}), we can identify this new exponent with the
spectral exponent $\zeta_s(d=1,\anhexp=2)\approx 1.39$.

\subsubsection{Critical exponents as a function of $\anhexp$}
\begin{figure}[tb]
\includegraphics[width=\columnwidth]{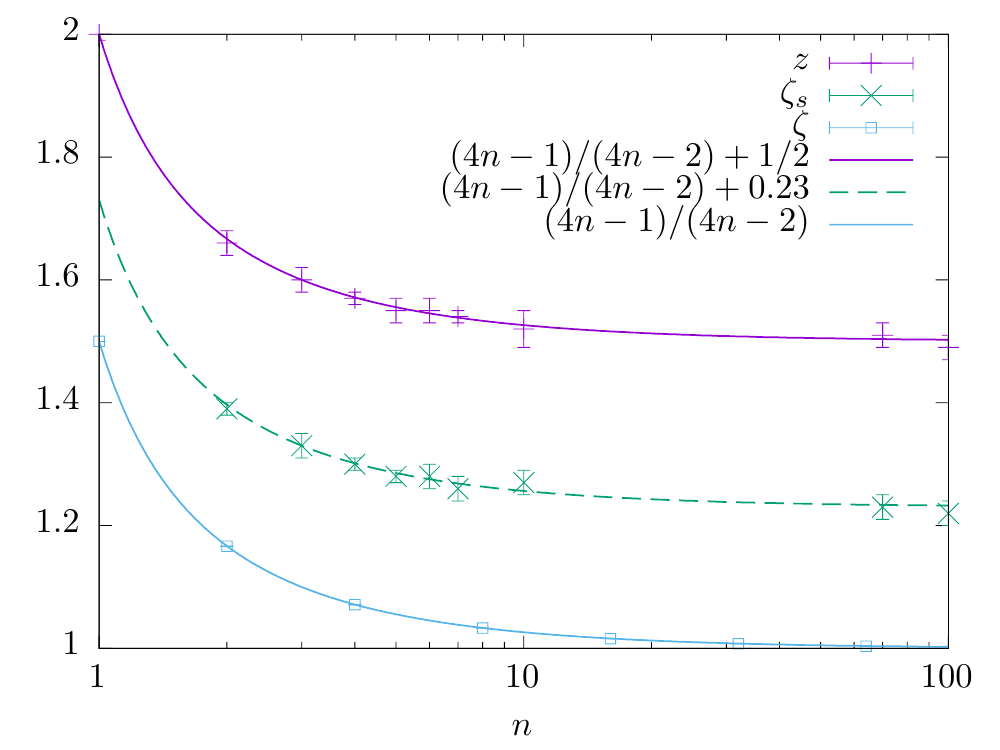}
\caption{ Global roughness exponent $\zeta$, spectral roughness exponent
$\zeta_s$, and dynamical exponent $z$  obtained from simulations of the $d=1$
harmonic and anharmonic Larkin models as a function of the elasticity exponent
$\anhexp$. $\anhexp=1$ corresponds to the usual harmonic Larkin Model, while
$\anhexp>1$ correspond to different anharmonic Larkin models. The spectral
roughness exponent equals the global exponent only for the harmonic case,
$\anhexp=1$. For $n>1$, we find $\zeta_s(d=1,n) - \zeta(d=1,n) \approx 0.23$, as
indicated by the dashed line.}
\label{fig:exp_vs_alpha}
\end{figure}

Repeating the above procedure, we have  fitted $z$, $\zeta$, and $\zeta_s$ for
different values of $\anhexp$, from $\anhexp=2$ to $\anhexp=100$. In
Fig.~\ref{fig:exp_vs_alpha} we show the corresponding results, where it can be
appreciated that numerical simulations perfectly agree, within the error
bars~\footnote{We use different criteria to estimate the error bars for each
exponent. In the case of $\zeta$, we estimate the uncertainty fitting the data
of Fig.~\ref{fig:w2_vs_L}. The error bar corresponding to $\zeta_s$ is obtained
by fitting the slope in the rescaled structure factors, shown in
Fig.~\ref{fig:sq_scaled_alpha_2}. For $z$ we make a conservative estimation of
its uncertainty by determining the range where a good collapse can be seen.},
with the predicted exponents $\zeta(d=1,\anhexp)$ and $z(d=1,\anhexp)$ (shown by
solid lines), from Eqs.~(\ref{eq:prediction_zeta}) and~(\ref{eq:prediction_z}).
In all cases, the fitted spectral exponent is found to be $\zeta_s(d=1,\anhexp)
> \zeta(d=1,\anhexp)$. We also observe that all exponents tend to converge to a
well defined finite value in the large $\anhexp$ limit.  The results also show a
roughly constant difference, $\zeta_s(d=1,\anhexp)-\zeta(d=1,\anhexp) \approx
0.23$ for $\anhexp>1$, i.e., for all the anharmonic cases. This is very
interesting, as we do not have any analytical prediction for
$\zeta_s(d=1,\anhexp)$.

\section{Discussions}
\label{sec:discussions}
We find an excellent agreement of the general scaling predictions of
Sec.~\ref{sec:scalingarguments} with static and dynamical numerical simulations
of a $d=1$ interface as a function of the anharmonicity parameter $\anhexp$.
These predictions include the global roughness exponent $\zeta$, the dynamical
exponent $z$, and the harmonic to {anharmonic} crossover length
$L_{\text{anh}}$. We also obtain {numerically} the spectral roughness exponent
$\zeta_s$, for which no analytical prediction is available, and we find that it
is a different exponent, at least for $d=1$ and $n>1$. In the following sections
we discuss possible connections of the ALM with other problems, and then
we discuss the anomalous scaling.

\subsection{Family of Brownian functionals}
\label{eq:brownian}
It is interesting to note that the static solution $u_L$ of
Eq.~(\ref{eq:staticequation}) is directly related to a family of Brownian
functionals~\cite{majumdar2006} parametrized by $\anhexp$,
\begin{equation}
u_L = -\int_0^L dx\; \text{sign}[{B(x)}] |B(x)|^{\frac{1}{2\anhexp-1}},
\label{eq:brownian_func}
\end{equation}
with the Brownian path $B(x)$ given by
\begin{equation}
B(x)=\int_0^L dx\; f(x).
\label{eq:rw}
\end{equation}
From the statistical properties assumed for the elementary pinning forces
[Eq.~(\ref{eq:pinningforceprops})], $B(x)$ is clearly a Wiener process.

Let us now analyze the family of Brownian functionals. The $\anhexp=1$ case,
\begin{equation}
u_L =\int_0^L dx\; B(x),
\end{equation}
simply corresponds to the random-acceleration process (see, for instance,
Refs.~\cite{burkhardt2007,hilhorst2008,majumdar2010}). This well-known
process has a Gaussian end-point distribution with a particular super-extensive
variance,
\begin{eqnarray}
P(u_L,\anhexp=1) &=& \frac{\exp[-u_L^2/2\sigma_L^2]}{\sqrt{2\pi \sigma_L^2}},
\nonumber \\
\sigma_L^2 &\sim& L^3.
\label{eq:randomacc}
\end{eqnarray}
Hence, $u_L \sim L^{3/2}$ is in perfect agreement with the scaling result $u_L
\sim L^{\zeta}$, with the global roughness exponent
\mbox{$\zeta=\zeta(d=1,\anhexp=1)=3/2$} predicted in
Eq.~(\ref{eq:prediction_zeta}).

On the other extreme of the ALM family, when $\anhexp \to \infty$ we get
\begin{equation}
u_L =\int dx\; \text{sign}[B(x)],
\end{equation}
which corresponds to the so-called arcsine-law L\'{e}vy process~\cite{Levy1940}.
We have also an exact result for this case,
\begin{equation}
P(u_L, \anhexp=\infty)=\frac{1}{\pi} \frac{1}{\sqrt{(L-u_L)(u_L+L)}}.
\label{eq:levy}
\end{equation}
Its variance grows as $\sim L^2$ implying that \mbox{$u_L \sim L$}, which is
also in perfect agreement with our predicted
\mbox{$\zeta(d=1,\anhexp=\infty)=1$} in Eq.~(\ref{eq:prediction_zeta}).

\begin{figure}[tb]
\includegraphics[width=\columnwidth]{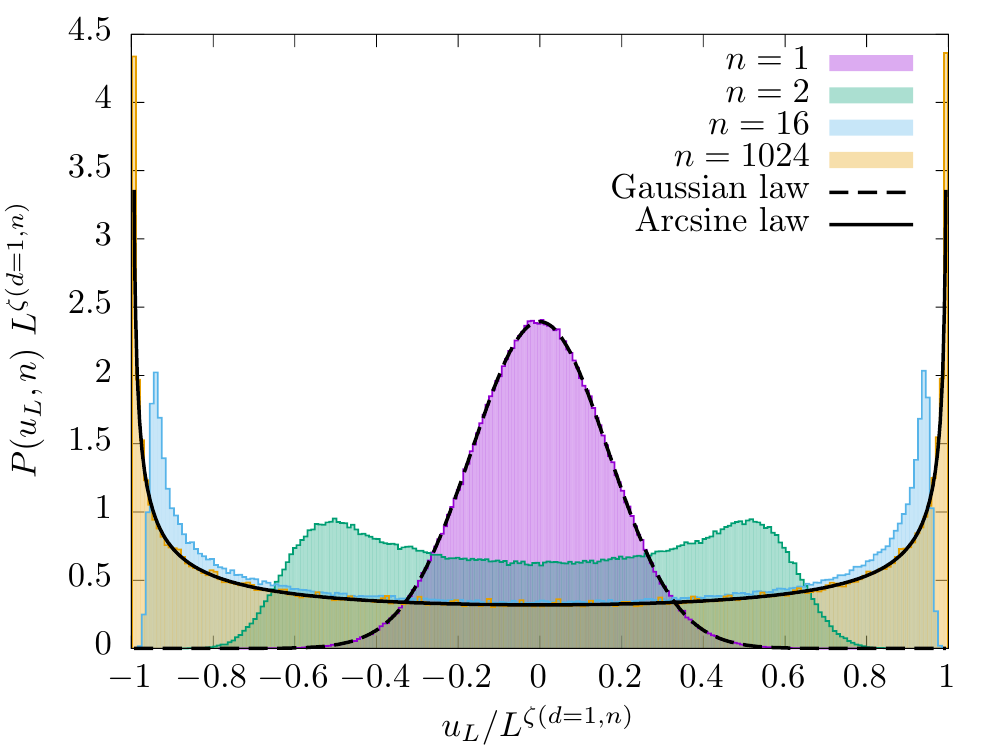}
\caption{Rescaled end-point distribution $P(u_L,\anhexp)$. The dashed and the
solid lines indicate the analytical results known for the random acceleration
process, and for the L\'evy arcsine-law process, respectively. The first maps to
the harmonic Larkin model ($\anhexp=1$), the second maps to the extreme
($\anhexp=\infty$) anharmonic Larkin model.}
\label{fig:conf_modelo}
\end{figure}

For intermediate cases, $1<\anhexp<\infty$, $P(u_L,\anhexp)$ is, to the best our
knowledge, not known analytically. In Fig.~\ref{fig:conf_modelo} we show
$P(u_L,\anhexp)$ obtained numerically from Eq.~(\ref{eq:conf_JLI}), for several
values of $\anhexp$. These end-point distributions  characterize the family of
Brownian functionals describing the $d=1$ ALM. We observe that, in general, all
$\anhexp>1$ distributions are symmetric but non-Gaussian, with twin peaks far
from the origin that grow with increasing $\anhexp$, and diverge when $\anhexp
\to \infty$, as the L\'evy arcsine process is approached.

Different methods, such as the Feynman-Kac formula (see, for instance,
Ref.~\cite{majumdar2006}), could be used in principle to get
$P(u_L;\anhexp)$ analytically for general $\anhexp$, and to compare with our
numerical results of Fig.~\ref{fig:conf_modelo}.

\subsection{Relation to variants of the KPZ equation}
Following the connection made in Ref.~\cite{Rosso2001b}, between the QKPZ
equation and the QEW equation with anharmonic elasticity at depinning, one may
ask whether the ALM roughness properties can be related to other nonlinear
roughening equations. At this respect, it is worth noting that the KPZ equation
with temporally correlated noise,
\begin{equation}
\partial_t u = c_2 \partial_x^2 u + \lambda (\partial_x u)^2 + \eta(x,t),
\label{eq:kpzcorrelated}
\end{equation}
with the noise field such that $\langle \eta(x,t) \rangle=0$, and $\langle
\eta(x,t)\eta(x',t') \rangle = |x-x'|^{2\psi-d}\,|t-t'|^{2\phi-1}$, was found to
satisfy the scaling relation~\cite{barabasi},
\begin{equation}
z(1+2\phi)-2\zeta-d+2\psi=0.
\label{eq:relationkpzcorrelated}
\end{equation}
Interestingly, for quenched spatially uncorrelated
noise~\cite{cates1988statistics} ($\phi=1/2$ and $\psi=0$), we get
$2z-2\zeta-d=0$, which is identical to Eq.~(\ref{eq:prediction_z}). This can be
related to the fact that Eq.~(\ref{eq:relationkpzcorrelated}) is derived from
the nonrenormalization of the temporal component of the noise correlator, while
Eq.~(\ref{eq:prediction_z}) was derived assuming, analogously, the no-scaling of
the random force amplitude. Since the global roughness and dynamic exponents of
the ALM are parametrized by $n$, however, one can ask whether a value of $n$
exists such that the ALM and the quenched noise KPZ share the same exponents
$\zeta$, $\zeta_s$, and $z$.

\begin{figure}[tb]
\includegraphics[width=\columnwidth]{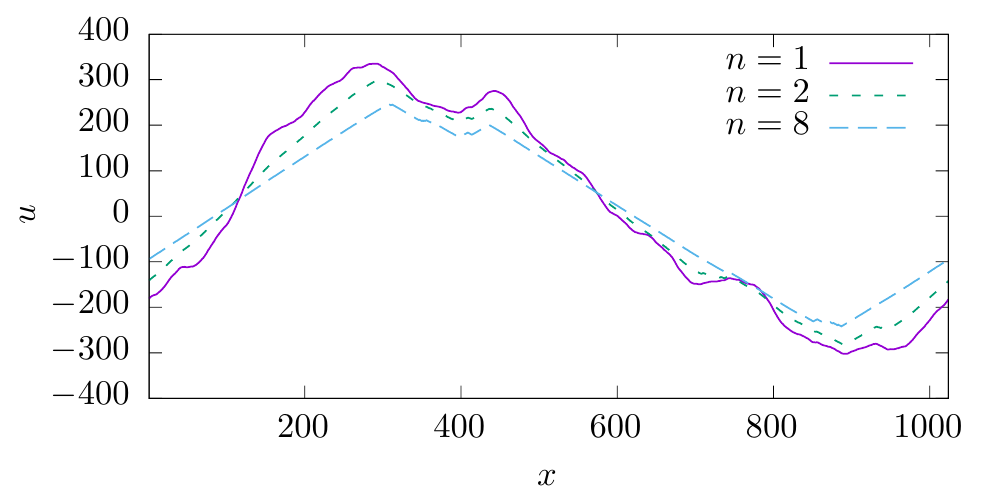}
\caption{Snapshots of $L=1024$ interfaces using the dynamic solution for
different $n$ and the same realization of the random forces $f(x)$. The $n=1$
corresponds to the $1d$ harmonic Larkin model. The interfaces become more
``faceted'' with further increasing $n$.}
\label{fig:faceted}
\end{figure}

At this respect, numerical simulations of temporally correlated KPZ for $d=1$
and $\lambda>0$ report, for the  $\phi=1/2$ quenched-noise limit, {$\zeta
\approx 1.06$, $z \approx 1.54$} \cite{song2016} and  {$\zeta \approx 1.07$, $z
\approx 1.15$} \cite{szendro2007}. Exponents are thus roughly consistent with
the $n \approx 4$--$5$ ALM, where $\zeta \approx 1.06 \pm 0.01$ and $z \approx
1.56$. Interestingly, in Ref.~\cite{szendro2007} it is shown that configurations
display facets, with an spectral exponent $\zeta_s \approx 1.5$, which is
slightly larger than the $\zeta_s \approx 1.29 \pm 0.01$ we find for $n \approx
4$--$5$ ALM. More recent results for the KPZ with temporally correlated noise
approaching the limit $\phi=1/2$, report $\zeta_s \approx 1.3$, showing a closer
agreement with the $n \approx 4$--$5$ universality class of the
ALM~\cite{aleslopez2019}. The quenched noise KPZ with $\lambda<0$, on the other 
hand, has been associated (though for more general random forces) to faceted
interfaces~\cite{jeong1999}, similar to the ones we show for $d=1$ in
Fig.~\ref{fig:faceted}. The possible connection of the ALM for general $n$ with
$n$-dependent extensions of Eq.~(\ref{eq:kpzcorrelated}) remains an interesting
open question.

Finally, it is also worth pointing out that the $d=1$ connection of the ALM with
Brownian functionals of the previous section also suggests, from a different
perspective, a connection with other KPZ variants. The Feynman-Kac formula is
indeed connected to a family of quantum Hamiltonians describing a single
particle in a family of potentials $pU(X)$ with $p$ a parameter (see, for
instance, Ref.~\cite{majumdar2006}). The quantum propagator satisfying each
imaginary-time Schr{\"o}dinger-like equation can then be mapped, by a suitable
``Cole-Hopf'' transformation, to a particular forced KPZ
equation~\cite{kardar_2007}. From Eq.~(\ref{eq:brownian_func}) we have
$U(X)=\text{sign}[{X}] |X|^{\frac{1}{2\anhexp-1}}$ so the potential is
parametrized by both $p$ and $\anhexp$. It would be interesting to test these
connections further.

\subsection{Two different roughness exponents}
The ALM displays interesting anomalous scaling properties. In particular, in the
steady state, the spectral roughness exponent does not coincide with the global
roughness exponent ($\zeta_s>\zeta$ for $\anhexp>1$) in $d=1$. This is in sharp
contrast to what happens for the LM and many other linear surface growth models,
such as the EW equation, where interfaces are characterized by a single
roughness exponent, and the usual Family-Vicsek scaling applies~\cite{barabasi}.
It is worth noting, however, that also nonlinear models such as the KPZ equation
with additive noise~\cite{bustingorry_kpz_2007}, or the zero temperature QEW
equation at the depinning threshold~\cite{Ferrero_2013}, seem to display a
normal Family-Vicsek scaling of the two point correlations functions, such that
$\zeta \approx \zeta_s$, with $\zeta_s \approx 1/2$ in the former and $\zeta_s
\approx 1.25$ in the latter, in $d=1$. Therefore, nonlinearity, either in the
elasticity or in the disorder, is not sufficient to produce the anomaly, even
combining it with super-rough behavior $\zeta_s>1$, as it occurs in the QEW at
depinning. We also note that breaking the tilt symmetry (as for instance in the
KPZ equation), or breaking the statistical tilt symmetry (as for instance in the
$d=1$ QKPZ equation at depinning where $\zeta_s \approx \zeta \approx
0.63$~\cite{Rosso2001b,kolton2009}), does not seem to be sufficient for
producing the anomaly either.

However, other models are known to display the $\zeta \neq \zeta_s$ anomalous
behavior. The anomalous case with $\zeta_s<1$, so called ``intrinsic,'' has
been, for instance, observed~\cite{intrinsic_example_1997}. The case
$\zeta_s>1$ has been observed, however, in the Sneppen model A of self organized
criticality~\cite{PhysRevLett.84.2199}. Interestingly, in Ref.
\cite{PhysRevLett.84.2199}, the $\zeta \neq \zeta_s >1$ behavior has been
associated to generic ``faceted'' stationary interfaces, and a simple model of
random $\pm 1$ slopes used to illustrate the association. In
Fig.~\ref{fig:faceted}, we show that this is indeed qualitatively the case for
the ALM family parameterized by $\anhexp$ in $d=1$. As far as we can tell, the
generic  origin of the $\zeta \neq \zeta_s$ behavior, either in the $\zeta_s<1$
(``intrinsic'') or $\zeta_s>1$ (``faceted'') remains open, however.

We speculate that the observed anomalous roughness property $\zeta \neq \zeta_s$
of the $d=1$ ALM should be related to the shape of $P(u_L,\anhexp)$ for
$\anhexp>1$. Indeed, only for the LM ($\anhexp=1$) the distribution is Gaussian
and $\zeta=\zeta_s$. For $\anhexp>1$ it is non-Gaussian and presents two twin
peaks and, as expected, the larger and sharper they are, the more faceted the
individual configurations look like. Since the non-Gaussian bimodal shape of
$P(u_L,\anhexp)$ presents  scale invariance, the associated faceted structure
should be also scale invariant and thus visible in the structure factor shape.
This makes us conjecture that $\zeta_s(d=1,\anhexp)$ must be somehow coded in
$P(u_L,\anhexp)$. It would be useful to test this hypothesis for $d=1$ by
analytical calculations using, for instance, Brownian functionals techniques.

For $d>1$, however, it would be useful to go beyond heuristic methods and use
the Gaussian variational method (see, for instance, Ref.~\cite{AGORITSAS2012}),
or renormalization group techniques, to calculate both $\zeta$ and particularly
$\zeta_s$, which we were unable to estimate by simple scaling arguments. One
important question is when $\zeta_s \neq \zeta$, and to know if $\zeta_s$ is
universal. In other words, the question is whether $\zeta_s$ is needed to make a
finer classification into (anomalous) universality classes.

\section{Conclusions}
\label{sec:conclusion}
We have studied the roughness and dynamical properties of the anharmonic Larkin
model. By Flory-like scaling arguments, we have obtained the global $\zeta$
roughness exponent, the dynamic exponent $z$, and the harmonic to anharmonic
crossover length scale for arbitrary both $d$ and $n$. An excellent agreement is
obtained by comparing with numerical calculations in $d=1$, and we have showed
that, in this case, the model directly relates to a family of Brownian
functionals parameterized by $n$; ranging from the random-acceleration model
($n=1$) to the L\'evy arcsine-law problem ($n = \infty$).  It would be
interesting to exploit this connection, not only to find analytical methods of
solution, but to link or map the $d=1$ ALM model to other problems.

From the analytical and numerical calculations, we have found an intriguing
anomalous scaling for the $d=1$ case. The spectral roughness exponent $\zeta_s$
is different from the global $\zeta$, and $\zeta_s > \zeta >1$, for $n>1$. On
the one hand, this implies that the small gradient expansion for the elastic
energy is always compromised in the thermodynamic limit, even in the  $n \gg 1$
hard-constraint limit. On the other hand, that we need two roughness exponents,
already at the level of the two point correlation functions, to describe the
self-affine geometry of the interface. It would be interesting to see if the
``faceted'' scaling anomaly $\zeta_s \neq \zeta$ persists at larger dimensions,
and in particular, to find an analytical estimate of $\zeta_s$ as a function of
dimension $d$ and the anharmonicity parameter $n$ using, for instance,
variational or renormalization group approaches. This would allows us to test
the universality of $\zeta_s$, and to point out the mechanism originating the
anomaly in general cases.

Finally, although the Larkin model is usually a local approximation for more
realistic models of disordered elastic systems with many metastable states, it
would be nevertheless interesting to check some of our results experimentally.
This is in principle  possible in systems with anisotropically correlated random
forces. Indeed, interesting anomalous scaling was found for interfaces with
``columnar noise,'' both theoretically \cite{szendro2007} and experimentally
\cite{soriano2002}. Finite systems with either stiff elastic couplings or very
weak disorder can also comply with the Larkin approximation for disorder at the
relevant scales. Larkin random forces could be also spontaneously generated, by
coarse-graining, well beyond the Larkin length in large driven periodic systems
such as elastic chains~\cite{Bustingorry2010}. In all these cases, the
additional anharmonicity needed for realizing the ALM described here may arise
from some nonlinear local interaction breaking the tilt symmetry of the clean
(i.e. nondisordered) system. Directed polymers or membranes in a layered
Matteron-de Marsily scalar flow field ~\cite{oshanin1994} may be a realization
of the ALM schematically represented in Fig.~\ref{fig:esquemaalm}.

\begin{acknowledgments}
We thank S. Bustingorry, A. Rosso,  C. Texier, and T. Giamarchi for useful and
motivating discussions. This work was partly supported by Grants No.
PICT2016-0069/FONCyT and No. UNCuyo C017, from Argentina. We have used Mendieta
Cluster from CCAD-UNC and GPGPU-CAB Cluster from CAB, which are part of
SNCAD-MinCyT, Argentina.
\end{acknowledgments}

\appendix
\section{Larkin length with pure non-harmonic elasticity}
\label{sec:anharmoniclarkinlength}
If we see Eq.~(\ref{eq:general_differential}) as a short-scale approximation of
the full model of Eq.~(\ref{eq:motion}) with
Eq.~(\ref{eq:nonharmonicelasticity}), then it is useful to get the corresponding
(``anharmonic'') Larkin length.

By balancing the elastic force with the pinning force of a piece of linear size
$l$, we get $c_{2\anhexp} u^{2\anhexp-1} l^{-2(2\anhexp-1)-1} \approx \ampdis
l^{-d/2}$. We can define $L_c(d,\anhexp)$ as the length $l$ corresponding to
$u=\xi$,
\begin{equation}
\xi = \left({\frac{\ampdis}{c_{2\anhexp}}}\right)^{\frac{1}{2\anhexp-1}}
      L_c^{\zeta(d,\anhexp)},
\end{equation}
so we obtain
\begin{equation}
L_c(d,\anhexp) = \left[\xi^{2\anhexp-1}
\left(\frac{c_{2\anhexp}}{\ampdis}\right)\right]^{2/(4\anhexp-d)}.
\end{equation}
Thus, for $\anhexp=1$, we recover the well-known harmonic result
$L_c(d,1)=\left({\xi\ c_{2}/\ampdis}\right)^{2/(4-d)}$. For $\anhexp=\infty$,
however, $L_c \to \xi$. The length $L_c(d,n)$ marks the crossover to the
random-manifold regime of the purely anharmonic interface. While the Larkin
regime is described by the exponents $\zeta(d,n)$
[Eq.~(\ref{eq:prediction_zeta})] and $z(d,n)$ [Eq.~(\ref{eq:prediction_z})],
regardless of the presence of the driving force $F$, the random-manifold
exponents do change with $F$. The equilibrium random manifold exponents (present
at equilibrium or in the creep regime at intermediate
scales~\cite{kolton2006b,kolton2009}) are expected to be the same as those for
the harmonic elasticity or QEW equation, while the depinning random manifold
exponents (present at depinning or in the creep regime at large scales for
vanishing velocities ~\cite{kolton2006b,Kolton2009b}) coincide with the ones of
the QKPZ equation~\cite{Rosso2001b}.

\bibliography{references.bib}
\end{document}